\documentstyle[12pt]{article}
\def\V{{\cal V}}
\def\H{{\cal H}}

\newcommand{\be}{\begin{eqnarray}}
\newcommand{\en}{\end{eqnarray}}

\begin{document}
\begin{titlepage}
\begin{flushright}
EFI 95-54 \\
MPI-Ph/95-82  \\
\end{flushright}

\begin{center}
\vskip 0.3truein

{\bf {Dispersion Relations}}
\vskip0.10truein

{\bf {in Gauge Theories with Confinement}}
\footnote{Plenary talk presented at the XVIIIth International
   Workshop on High Energy Physics and Field Theory,
   Moscow-Protvino, June 1995. To be published in the Proceedings. }
\vskip0.5truein

{Reinhard Oehme}
\vskip0.2truein

{\it Enrico Fermi Institute and Department of Physics}

{\it University of Chicago, Chicago, Illinois, 60637, USA}
\footnote{Permanent Address}

{\it and}

{\it Max-Planck-Institut f\"{u}r Physik}

{\it - Werner-Heisenberg-Institut -}

{\it 80805 Munich, Germany}
\end{center}
\vskip0.2truein
\centerline{\bf Abstract}
\vskip0.13truein

The analytic structure of {\it physical} amplitudes
is considered for gauge theories with confinement
of excitations corresponding to the elementary fields.
Confinement is defined in terms of the BRST algebra.
BRST-invariant, local, composite fields are introduced,
which interpolate between physical asymptotic states.
It is shown that the singularities of physical
amplitudes are the same as in an effective theory
with only physical fields. In particular,
there are no structure singularities (anomalous thresholds)
associated with confined constituents, like quarks and gluons.
The old proofs of dispersion relations for
hadronic amplitudes remain valid in QCD.

\end{titlepage}
\newpage
\baselineskip 20 pt
\pagestyle{plain}

%\centerline{\bf I. INTRODUCTION}
\vskip0.2truein
It is the purpose of this talk, to give a survey
of the problems involved in the derivation of analytic
properties of {\it physical} amplitudes in gauge theories
with confinement. This report is restricted to a brief
resum\'{e} of the essential points discussed in the talk.
\footnote{For more detailed discussions, I refer to
the report covering my talk at the 1994 ICMP in Paris
\cite{PAR}, and to the articles \cite{OPN}.}

Dispersion relations for amplitudes describing
reactions between hadrons, and for form factors
describing the structure of particles,
have long played an important r\^{o}le in particle
physics \cite{GMO, RON, GNO, GOR}.
Analytic properties of Green's functions
are fundamental for proving many important results in
quantum field theory. With physical amplitudes being
boundary values of holomorphic functions of several
complex variables, appropriate different amplitudes can
be associated with the same analytic function.

In the past, analytic properties have been studied within the
framework of a local field theory of hadrons, formulated
in a state space of definite metric
\cite{BOT, JLD, BOG, DSF}. Heisenberg fields
associated with hadrons were introduced. They interpolate
between asymptotic states of non-interacting physical
particles, and they commute or anti-commute at space-like
separations. For the Fourier transforms of retarded and
advanced products of these operators, the local commutativity
gives rise to tubes (wedges) $W^{\pm}$ of holomorphy, while the
boundary values at real point are tempered distributions. As a
consequence of spectral conditions resulting from lower
bounds for the spectrum of the energy-momentum operator,
one obtains real domains $R$, where retarded and advanced
amplitudes coincide as distributions. Under these conditions,
the {\it Edge of the Wedge Theorem} \cite{BOT} is applicable. It
provides an analytic function in the union of the
wedges $W = W^{+} \cup W^{-} $ and a finite, complex
neighborhood $N(R)$ of the real domain $R$ :
$W \cup N(R) $. The task is then to construct
the {\it Envelope of Holomorphy} $E(W \cup N(R) ) $ of
this basic region of analyticity \cite{BRO, BOT, BOG}.
The envelope is the
largest domain into which all functions, which are regular
in the basic region $W \cup N(R) $,
can be continued. It is characteristic
for the input, and additional information is needed
in order to get larger regions of analyticity. More detailed
use of unitarity, would be required,
and of specific features of the spectrum,
but these conditions are difficult to
implement \cite{MAT}.

The limitations of general derivations can usually be
understood as being related to singularities which describe
structure due to unphysical constituents of the physical
particles involved in the amplitude (unphysical anomalous
thresholds). The simple spectral conditions used are
not sufficient to eliminate these constituents.

It is sometimes helpful to turn the problem around, and
ask for singularities of an amplitude, which are definitely
expected on the basis of the spectrum of the theory.
Here weak coupling perturbation expansions of the effective
hadronic theory can be very useful.
Even though these expansions may not provide
a reasonable approximation of the amplitude itself, they
can be a guide to the {\it location} of singularities
\cite{MAN}.

There are many details involved in the derivation of
dispersion relations and related analytic properties of
Green's functions \cite{BOT, BOG, SYM, LEH, OTG},
but the path via the Edge of the Wedge
Theorem, and the Envelope of Holomorphy, is the essence of
the problem. Although the theory of functions with several complex
variables is the natural framework for obtaining regions
of analyticity, in special cases, like those involving
only one complex four-vector, methods from the theory of
differential equations and of distributions can be used
in order to obtain the analytic domain corresponding to
the envelope discussed above \cite{BOT, BOG}.

\vskip0.2truein
In this talk, we are interested in the analytic structure
of physical amplitudes in gauge theories with confinement.
We will often use the language of QCD. In these theories,
the spectrum is not related to the elementary Heisenberg
fields appearing in the initial formulation. Rather, the
elementary fields correspond to unphysical, confined
excitations. In the effective hadron theories discussed
above, the derivation of dispersion relations, and of other
analytic properties, is quite rigorous, given the basic
axioms of quantum field theory in a state space with
definite metric. In contrast, for gauge theories with
confinement, we will have to use several features of the
theory, which have not been proven rigorously in the
non-perturbative framework. The required assumptions will be
discussed in the following. They are mainly concerned
with the definition of confinement with the help of the
BRST-algebra, and with the construction of local, BRST-
invariant physical fields as composites of confined
Heisenberg fields.

Since we require a covariant formulation
of the theory, we must use a quantization in a state
space $\V$ of indefinite metric \cite{KOJ}. This space contains
quanta like ghosts and longitudinal and space-like gluons,
which are unphysical even in the weak coupling limit.
We use the BRST-algebra \cite{BRS} in order to define an invariant,
physical state space $\H $ with positive definite metric as a
cohomology of a nilpotent BRST-operator $Q$. The assumptions
involved here are the non-perturbative existence of
a BRST-operator, and its completeness \cite{SPI, ROC}. The latter
notion implies that all states $\Psi \in \V $, which satisfy
$Q\Psi = 0 $, and which have zero norm, are of the form
$\Psi = Q\Phi, ~~\Phi \in \V $. This means, that states
with zero ghost number, containing ghost-antighost pairs,
are eliminated. They have indefinite metric and would
make it impossible to define a physical state space
with definite norm. There are arguments for completeness,
but I do not know of a general proof in four-dimensional
gauge theories like QCD. In certain string theories,
completeness has been proven explicitly, but these are
more simple structures.

Already in weak-coupling perturbation theory, the ghosts
and the transverse and time-like gluons
are eliminated from the physical state space $\H $ in a
kinematical fashion. They are not color singlet states,
but form quartet representations of the BRST-algebra.
In the full theory, we expect that also quarks and
transverse gluons are confined, and do not appear as
elements of $\H $, at least at zero temperature. They
also form quartet representations, together with
other unphysical states. With certain limitations
concerning the number of flavors (less than ten for
QCD), we have given arguments that, for dynamical reasons,
transverse gluons cannot be elements of the cohomology
space $\H $ \cite{ROC}.
Some more preliminary methods also exclude quarks
\cite{NIC}. Our arguments for confinement are
based upon superconvergence relations for the gluon
propagator \cite{WZS, ROS}, and they involve
renormalization group methods. These arguments are
valid for zero temperature. At finite temperatures,
a new dimensionful parameter is present, and there
may be de-confinement. If our methods are applied to
certain $N=1 $ SUSY models \cite{RSU}, as far as the
number of flavors is concerned, they agree with results
obtained on the basis of duality and holomorphy of the
superpotential \cite{SEI}. An approximately linear
quark-antiquark potential is obtained on the basis
of superconvergence, with the same restrictions for
the number of flavors \cite{RLP, NLP}.

Confinement of quarks and gluons does not necessarily
imply the existence of massive states in $\H$, which can
be interpreted as hadrons. But, possibly with further
restrictions of the number of flavors, the existence of
hadrons may be a reasonable assumption. $N=1$ SUSY models
are encouraging in this respect. For our purpose, we
assume that the BRST singlet states, which span $\H $,
are hadrons.

As we have mentioned, local, interpolating Heisenberg
fields are the basis for obtaining analytic properties.
The support of retarded and advanced amplitudes implies
Fourier transforms, which are analytic in wedges containing
no real points. In QCD, we need to construct local,
composite operators, which
interpolate between non-interacting, asymptotic hadron
fields. These local Heisenberg fields of hadrons must be
BRST invariant operators, constructed from the elementary
fields associated with confined quarks and gluons. The
construction of local composite operators has been studied
extensively in quantum field theory \cite{WZC}, in
particular as leading terms of operator product expansions
\cite{WLS}. These have been derived in renormalized, weak
coupling perturbation theory, but here we use the
composite fields in a non-perturbative framework. It
is important to realize, that the local character of these
fields is related to the center-of-mass motion of the
constituents, and does {it not} imply a point-like
structure of the bound system. In quantum field theory,
the extended distribution of a particle, viewed as
a composite of other particles, is described by the
anomalous thresholds (structure singularities) of form
factors, scattering amplitudes, and other Green's functions
\cite{RAN, NAM, KSW, RHF}. These singularities are
particularly prominent for loosely bound systems, like
the deuteron, for example, where the range of the wave
function is much larger than the size of the pion cloud.

We find, that local, composite fields are a common
feature of causal field theories. We assume that possible
embedings of QCD into more comprehensive schemes are
not important for confinement, and for scattering processes
well below the Planck mass. If local field theory is considered
as a low energy limit of string theory, we may perhaps
expect deviations from microscopic causality at very
small distances, and corresponding corrections to dispersion
relations \cite{DSF}.

The other important aspect of the composite hadron fields
is their BRST-invariance.
\footnote{See \cite{ADC} for a discussion of classical,
local and gauge invariant composite fields in QCD.
I would like to thank Professor Divakaran for bringing
this paper to my attention.} If applied to an invariant
ground state $\Psi_0 $ with $Q\Psi_0 = 0 $, these fields
generate states $\Psi $, which again satisfy $Q\Psi = 0 $. Hence
the states $\Psi $ are also representatives of physical
states. As has been described in \cite{PAR, OPN}, if we
consider matrix elements of products of BRST-invariant
Heisenberg fields between physical states, any decomposition
with respect to a complete set of intermediate states
in $\V $ requires only a subset of these states which
which form a complete set
in $\H $. These features are most important for the
unitarity of the $S$-matrix and for dispersion relations.
They imply that only hadronic states play a r\^{o}le as
absorptive thresholds in various channels of an hadronic
amplitude. It follows, that the spectral conditions
for hadronic amplitudes are the same as in the old,
effective theory. Our definition of confinement implies,
that in the collision of hadrons only hadrons are
produced as final states.

The construction of local, interpolating hadron fields
is not unique. There are equivalence classes of different
fields, which have the same asymptotic fields and
give rise to the same $S$-matrix. This is a consequence of
Borcher's theorem \cite{BOC}, which we use here in the physical
state space with positive definite metric, although it
can be generalized to indefinite metric spaces.

Some of the properties of hadronic amplitudes in QCD,
which we have discussed here, may appear to be
straightforward, once physical amplitudes are expressed in terms
of BRST-invariant, local operator fields. There are,
however, many subtile points due to the indefinite
metric of the full state space $\V $ \cite{SPI, KOO, ROC}.
There is no simple projection into the invariant space
$\H $. Unphysical states in $\V $
may well have components in the physical space,
or they may acquire such components after a Lorentz
transformation. An unphysical state is recognized
only by the fact that there exists an equivalence
transformation, which removes the component in the
physical space without changing the observables of
the theory. A detailed discussion of unitarity and
of the spectral conditions is therefore indicated
\cite{PAR, KOJ}.

The features of hadronic fields described above apply
to absorptive thresholds in a given channel of physical
amplitudes. It remains to discuss anomalous thresholds
or structure singularities, which have been described
earlier for the case of observable constituents. The
important question is, whether there are singularities
of hadronic amplitudes, which are related to the quark-gluon
structure of the hadrons. From the BRST-invariance of the
hadron fields, we infer implicitly, that such singularities
cannot be present, but a more explicit understanding is
desirable. We have shown in \cite{RHF}, that anomalous
thresholds are due to poles and absorptive branch points of
other hadronic amplitudes, which are related to the one
under consideration by analytic continuation into
appropriate lower Riemann sheets. In this way, in a
non-perturbative manner, we relate structure singularities
to absorptive thresholds, and these are only hadronic
if quarks and gluons are confined. As a consequence, there are
no anomalous thresholds associated with confined quarks and gluons.

We have already mentioned the example of the deuteron
as a composite system with loose binding and observable
constituents. The form factor of the deuteron is dominated
by anomalous thresholds well below the pion branch points.
The resulting distribution is as expected on the basis of the
Schr\"{o}dinger wave function. For hadrons, which may be
considered as loosely bound systems of heavy quarks, the
situation with respect to the quark-gluon structure is
completely different. Here the constituents are confined,
and there are no anomalous thresholds describing a long-
range quark structure, which one may expect from a
constituent quark model on the basis of the Schr\"{o}dinger
wave function. However, there is no problem in obtaining
a large mean-square radius with an appropriate form of the
discontinuities associated with hadronic thresholds
\cite{OPN, MEN, PAR}.
Where applicable, also hadronic anomalous thresholds may
contribute to an extended distribution.

\vskip0.7truein
\centerline{\bf ACKNOWLEDGMENTS}
\vskip0.2truein

It is a pleasure to thank Wolfhart Zimmermann, and the
Theory Group of the Max Planck Institut
f\"{u}r Physik, Werner Heisenberg Institut, for their
kind hospitality in M\"{u}nchen.

This work has been supported in part by the
National Science Foundation, grant PHY 91-23780.
\newpage
\vskip0.7truein
%centerline{\bf References}
%vskip0.2truein


\begin{thebibliography}{99}
\bibitem{PAR}
{R. Oehme, Analytic Structure of Amplitudes in Gauge
Theories with Confinement, Fermi Institute Report
EFI 94-39, hep-th/9412024; \\ Int. J. Mod. Phys.
{\bf A10} (1995) 1995-2401. }
\bibitem{OPN}
{R. Oehme, Mod. Phys. Lett. {\bf 8}, 1533 (1993);  \\
$\pi N$-Newsletter No. {\bf 7} (1992) 1. }
\bibitem{GMO}
{M. L. Goldberger, H. Miyazawa and R. Oehme, Phys Rev.
{\bf 99} (1956) 986. }
\bibitem{RON}
{R. Oehme, Phys. Rev. {\bf 100} (1955) 1503;
{\bf 101} (1956) 1174. }
\bibitem{GNO}
{M. L. Goldberger, Y. Nambu and R. Oehme, Ann. Phys.
(N.Y.)  {\bf 2} (1956) 226. }
\bibitem{GOR}
{M. L. Goldberger, Y. Nambu and R. Oehme, reported in
{\it Proceedings of the Sixth Annual Rochester Conference}
(Interscience, New York, 1956) pp. 1-7 ;
G. F. Chew, M. L. Goldberger, F. E. Low and Y. Nambu,
Phys. Rev. {\bf 106} (1957) 1337. }
\bibitem{BOT}
{H. J. Bremermann, R. Oehme and J. G. Taylor,
Phys. Rev. {\bf 109} (1958) 2178.}
\bibitem{JLD}
{R. Jost and H. Lehmann, Nuovo Cimento  {\bf 5},
1958 (1957);  \\
F. J. Dyson, Phys. Rev. {\bf 110} (1958) 1460. }
\bibitem{BOG}
{N. N. Bogoliubov, B. V. Medvedev and M. V. Polivanov,
{\it Voprossy Teorii Dispersionnykh Sootnoshenii}
(Fitmatgiz, Moscow, 1958); \\
N. N. Bogoliubov and D. V. Shirkov, {\it Introduction to
the Theory of Quantized Fields} (Interscience,
New York, 1959).}
\bibitem{DSF}
{R. Oehme, Nuovo Cimento {\bf 10} (1956) 1316; \\
R. Oehme, in {\it Quanta}, edited by P. Freund,
C. Goebel and Y. Nambu (University of Chicago Press,
Chicago, 1970) pp. 309-337. }
\bibitem{BRO}
{J. Bros, A. Messiah and R. Stora,
Journ. Math. Phys. {\bf 2} (1961) 639.}
\bibitem{MAT}
{A. Martin, in {\it Lecture Notes in Physics}
No. {\bf 3} (Springer Verlag, Berlin, 1969);
G. Sommer, Fortschritte der Physik, {\bf 18} (1970) 577;
and papers quoted in these articles. }
\bibitem{MAN}
{S. Mandelstam, Phys. Rev. {\bf 112} (1958) 1344. }
\bibitem{SYM}
{K. Symanzik, Phys. Rev. {\bf 100} (1957) 743. }
\bibitem{LEH}
{H. Lehmann, Suppl. Nuovo Cimento, {\bf 14} (1959) 1; \\
Nuovo Cimento {\bf 10} (1958) 1460. }
\bibitem{OTG}
{R. Oehme and J. G. Taylor, Phys. Rev. {\bf 113} (1959) 371.}
\bibitem{KOJ}
{T. Kugo and I. Ojima, Prog. Theor. Phys. Suppl.
{\bf 66} (1979) 1;
N. Nakanishi, Prog. Theor. Phys. {\bf 62} (1979) 1396;
K. Nishijima, Nucl. Phys. {\bf B238} (1984) 601;
I. B. Frenkel, H. Garland and G. J. Zuckerman,
Proc. Nat. Acad. Sci. USA, {\bf 83} (1986) 8442;
R. Oehme, Mod. Phys. Lett. {\bf A6} (1991) 3427;
N. Nakanishi and I. Ojima, {\it Covariant Operator
Formalism of Gauge Theories and Quantum Gravity}
(World Scientific, Singapore, 1990); F. Strocchi,
Comm. Math. Phys. {\bf 56} (1978) 57; Phys. Rev.
{\bf D17} (1978) 2010. }
\bibitem{BRS}
{C. Becchi, A. Rouet and R. Stora, Ann. Phys. (N.Y.)
{\bf 98} (1976) 287; \\ I. V. Tyutin, Lebedev Report
 No. FIAN 39 (1975) (unpublished). }
\bibitem{SPI}
{M. Spiegelglas, Nuc. Phys. {\bf B283} (1987) 205. }
\bibitem{ROC}
{R. Oehme, Phys. Rev. {\bf D42} (1990) 4209; ~
Phys. Lett. {\bf B155} (1987) 60. }
\bibitem{NIC}
{K. Nishijima, Prog. Theor. Phys. {\bf 75} (1986) 22;
K. Nishijima and Y. Okada, ibid. {\bf 72} (1984) 254;
K. Nishijima in {\it Symmetry in Nature}, Festschrift
for Luigi A. Radicati di Brozolo (Scuola Normale Superiore,
Pisa, 1989) pp. 627-655. }
\bibitem{WZS}
{R. Oehme and W. Zimmermann, Phys. Rev. {\bf D21} (1980) 474, 1661. }
\bibitem{ROS}
{R. Oehme, Phys. Lett. {\bf B252} (1990) 641.}
\bibitem{RSU}
{R. Oehme, ``Duality and Superconvergence'', (in preparation); \\
``Superconvergence, Confinement and Duality'', in {\it Proceedings
of the International Workshop on High Energy Physics }, Novy Svet, Crimea,
September 1995 (to be published), Fermi Insttute Report EFI 95-45.}
\bibitem{SEI}
{N. Seiberg, Phys. Rev. {\bf D49} (1994) 6857;
Nucl. Phys. {\bf B435} (1995) 129. }
\bibitem{RLP}
{R. Oehme, Phys. Lett. {\bf B232} (1989) 489. }
\bibitem{NLP}
{K. Nishijima, Prog. Theor. Phys. {\bf 77} (1987) 1053. }
\bibitem{WZC}
{W. Zimmermann, Nuovo Cimneto {\bf 10} (1958) 596;  \\
K. Nishijima, Phys. Rev. {\bf 111} (1958) 995.}
\bibitem{WLS}
{K. Wilson, Phys. Rev. {\bf 179} (1968) 1499;
K. Wilson and W. Zimmermann, Comm. Math. Phys. {\bf 24} (1972) 87;
W. Zimmermann, in {\it 1970 Brandeis Lectures}, edited by
S. Deser, M. Grisaru and H. Pendleton (MIT Press,
Cambridge, 1971) pp. 395-591;
W. Zimmermann, in {\it Wandering in the Fields}, edited by
K. Kawarbayashi and A. Ukawa (World Scientific,
Singapore, 1987) pp. 61-80.}
\bibitem{RAN}
{R. Oehme, Phys. Rev {\bf 111} (1958) 143;
Nuovo Cimento {\bf 13} (1959) 778. }
\bibitem{NAM}
{Y. Nambu, Nuovo Cimento {\bf 9} (1958) 610.}
\bibitem{KSW}
{R. Karplus, C. M. Sommerfield and F. H. Wichmann,
Phys. Rev. {\bf 111} (1958) 1187;
L. D. Landau, Nucl. Phys. {\bf B13} (1959) 181;
R. E. Cutkosky, J. Math. Phys. {\bf 1} (1960) 429. }
\bibitem{RHF}
{R. Oehme, in {\it Werner Heisenberg und die Physik
unserer Zeit}, edited by F. Bopp (Vieweg, Braunschweig,
1961) pp. 240-259;   \\
Phys. Rev. {\bf 121} (1961) 1840. }
\bibitem{ADC}
{M. Azam and P. P. Divakaran, Helv. Physica Acta {\bf 61}
(1988) 905. }
\bibitem{BOC}
{H.-J. Borchers, Nuovo Cimento {\bf 15} (1960) 784. }
\bibitem{KOO}
{Kugo, Ojima \cite{KOJ}.}
\bibitem{MEN}
{R. L. Jaffe and P. F. Mende, Nucl. Phys. {\bf B369} (1992) 189. }



\end{thebibliography}
\end{document}